\documentclass[manuscript]{aastex}
\usepackage{psfig,natbib,emulateapj5}
\shortauthors{Eracleous \& Halpern}
\shorttitle{Emission Lines of NGC~3065}

\slugcomment{To Appear in the Astrophysical Journal}

\def\ft#1{\tablenotemark{\,#1}}

\begin{document}
        
\def\aj{\rm{AJ}}                   
\def\araa{\rm{ARA\&A}}             
\def\apj{\rm {ApJ}}                
\def\apjl{\rm{ApJ}}                
\def\apjs{\rm{ApJS}}               
\def\apss{\rm{Ap\&SS}}             
\def\aap{\rm{A\&A}}                
\def\aapr{\rm{A\&A~Rev.}}          
\def\aaps{\rm{A\&AS}}              
\def\mnras{\rm{MNRAS}}             
\def\nat{\rm{Nature}}              
\def\pasj{\rm{PASJ}}    	   
\def\procspie{\rm{Proc.~SPIE}}     

\title{NGC~3065: A Certified LINER With Broad, Variable Balmer Lines}


\author{Michael Eracleous\altaffilmark{1}}
\affil{Department of Astronomy and Astrophysics, The Pennsylvania
State University, 525 Davey Lab, University Park, PA 16802}
\and 
\author{Jules P. Halpern\altaffilmark{1}} 
\affil{Department of Astronomy, Columbia University, 550 West 120th
St., New York, NY 10027}
\altaffiltext{1}{Visiting astronomer, Kitt Peak National Observatory,
which is operated by AURA, Inc., under a cooperative agreement with
the National Science Foundation}

\begin{abstract}

Motivated by the X-ray properties of the galaxy NGC~3065, we have
obtained new optical spectra which reveal that it has a low-ionization
nuclear emission-line region (LINER) as well as broad Balmer emission
lines, establishing it as an active galactic nucleus.  We also
examined an older spectrum from the CfA Redshift Survey which, lacking
broad Balmer lines, indicates that they appeared some time after 1980.
Thus NGC~3065 joins the set of LINERs with broad, variable Balmer
lines, which includes such well-known galaxies as NGC~1097 and
M81. Inspired by the sometimes double-peaked profiles of the variable
Balmer lines in other LINERs, we speculate that the broad Balmer lines
of NGC~3065 also come from an accretion disk. We illustrate the
plausibility of this hypothesis by fitting a disk model to the
observed H$\alpha$ profile. We also estimate the mass of the central
black hole as $(9\pm4) \times 10^7$~$M_{\odot}$ from the properties of
the host galaxy, which leads to the conclusion that the accretion rate
is only $\sim 2 \times 10^{-4}$ times the Eddington value, a property
that appears to be common among LINERs.  At such a low relative
accretion rate the inner accretion disk can turn into a
vertically-extended ion torus, which can illuminate the outer, thin
disk and power the broad-line emission. The reason for the sudden
appearance of broad Balmer lines is an open question, although we
suggest two possible explanations: tidal disruption of a star or a
sudden transition in the structure of the accretion disk.

\end{abstract}


\section{Introduction}

Low-ionization nuclear emission-line regions \citep[LINERs;][]{h80}
are a heterogeneous population of objects, with low-luminosity active
galactic nuclei (LLAGNs) and compact starbursts being two of the many
possible power sources \citep[see, for example][ for a
review]{f96}. From the point of view of the physics of accretion onto
compact objects, those LINERs that are true LLAGNs are particularly
interesting. At the low accretion rates onto the supermassive black
holes at the centers of these LINERs, the accretion flows are {\it
qualitatively} different from what occurs in more luminous Seyfert
galaxies and quasars.  It is therefore extremely interesting that in
the past few years several LINERs have been identified as {\it
bona-fide} AGNs based on the presence of broad, and sometimes
variable, Balmer lines in their optical spectra
\citep[e.g.,][]{hfsp97}. Here we report our discovery of a previously
unrecognized LINER and a true AGN with broad and variable Balmer lines
in the S0 galaxy NGC~3065 (also known as VII Zw 303).  Thus NGC~3065
joins the growing set of LINERs with variable, broad Balmer lines,
which includes such famous objects as NGC~1097 \citep*{sbw93} and M81
\citep{b96}.

NGC~3065 ($z=0.00667$, at a distance of 47.3~Mpc; from Tully 1987, but
for $H_0=50~{\rm km~s^{-1}~Mpc^{-1}}$\nocite{t87}) has been known to
have emission lines since the mid-1950s \citep*{hms56,bb65}. It was
detected as a radio source at 1.4~GHz in the NVSS Survey \citep{c98}
with a monochromatic luminosity of $1.2\times 10^{21}~{\rm
W~Hz^{-1}}$, and as an X-ray source with the {\it Einstein} IPC with
a 0.2--4~keV luminosity of $2.1\times 10^{41}~{\rm erg~s^{-1}}$
\citep*{fkt92}.  More recent X-ray observations with {\it ASCA} gave a
2--10~keV luminosity of $5\times 10^{41}~{\rm erg~s^{-1}}$, with a
spectrum that can be described either as a power-law with a photon
index of 1.8, or a 6~keV thermal plasma \citep{i98}. There was no
evidence for X-ray emission from a cooler thermal plasma as is found in many
other LINERs \citep[e.g.,][]{p99,t00}.  The hard X-ray spectrum
led \citet{i98} to suggest that NGC~3065 harbors a LLAGN and motivated
us to obtain new optical spectra in order to evaluate its credentials
further.

\section{Observations and Spectra: New and Old}

We obtained spectra of NGC~3065 with the MDM Observatory's 2.4m
telescope and Boller and Chivens CCD Spectrograph on 2000 May 31 UT,
and with the Kitt Peak National Observatory's 2.1m telescope and
GoldCam spectrograph on 2000 June 5 UT. In the former set of
observations we used a 1\farcs5 slit and a 150~mm$^{-1}$ grating with
a total exposure time of 2000~s to cover the wavelength range
3200--6860~\AA\ at a spectral resolution of 12.4~\AA. In the latter we
used a 1\farcs9 slit and a 600~mm$^{-1}$ grating with a total exposure
time of 1800~s to cover the wavelength range 5570--8540~\AA\ at a
spectral resolution of 4.2~\AA. In both cases, spectra were extracted
from a 1\farcs6 window along the slit. Because of the narrow apertures
used, which were comparable to the size of the seeing disk, the
absolute flux scale is somewhat uncertain. Wavelength calibration was
carried out with the help of arc spectra obtained immediately after
the object exposure, and flux calibration was carried out with the
help of standard stars observed on the same night and reduced in the
same manner as the object.  The final, reduced, and combined spectra
are shown in Figure~\ref{fig:spectra}. To investigate possible
variability of the {\it broad} H$\alpha$ line we also examined the
spectrum of NGC~3065 taken during the CfA Redshift Survey on 1980
February 10 with the Mt. Hopkins 1.5m telescope through a
3\arcsec$\times$12\arcsec\ aperture and originally reported by 
\citet{td81}. Although the flux scale of this spectrum was not
calibrated, it can still provide information on the presence and
profile of any {\it broad} H$\alpha$ line.

\hskip -0.3truein
\psfig{file=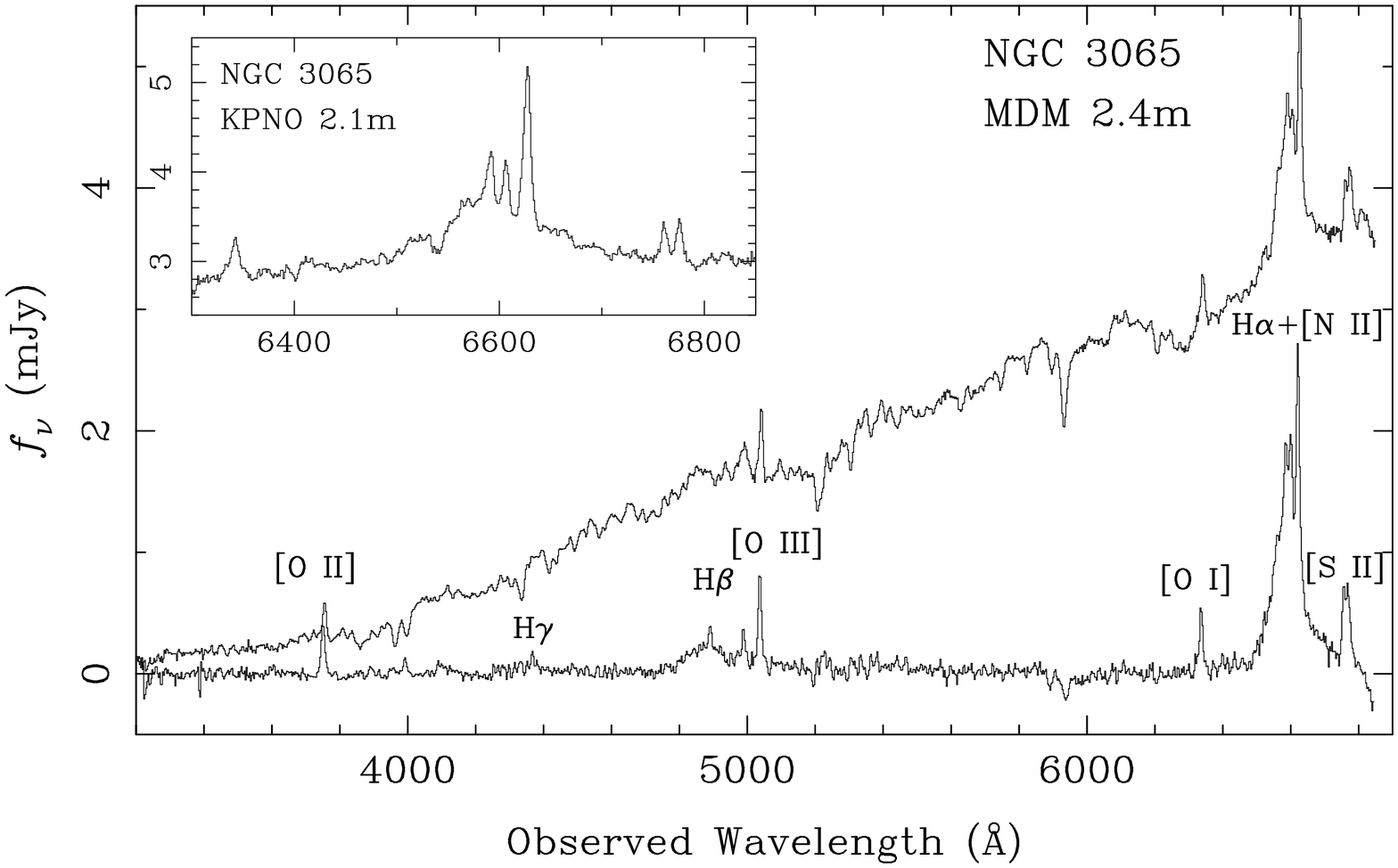,angle=-90,width=4in,rheight=2.5in}
\figcaption[figure1.ps] {Spectra of NGC~3065 obtained at MDM and KPNO
in 2000 June. The 3200--6860~\AA\ spectrum was obtained with the MDM
2.4m telescope and has a resolution of 12.4~\AA. The lower trace shows
the residual spectrum after subtracting the continuum (see \S3 of the
text for details) with the emission lines identified. The inset shows
the H$\alpha$ region of the higher-resolution (4.2~\AA) spectrum
obtained with the KPNO 2.1m telescope.
\label{fig:spectra} }

\section{Measured Spectroscopic Properties}

Although the optical continuum of NGC~3065 is dominated by starlight,
a broad H$\alpha$ line is obviously present (see
Figure~\ref{fig:spectra}). To isolate the emission lines and measure
their properties we modeled the continuum as a linear combination of
starlight and a non-stellar component and subtracted it. To describe
the starlight we experimented with spectra of elliptical or S0
galaxies (NGC~3379, NGC~4339, NGC 4365, and NGC~5322), while the
non-stellar continuum was assumed to have a power-law spectrum of the
from $f_{\nu}\propto \lambda^{\alpha}$. We found that the spectrum of
NGC~4339 provides an excellent match to the continuum of NGC~3065,
with no need for a non-stellar component. The non-stellar continuum
need not contribute more than 10\% of the flux at the wavelengths of
H$\alpha$ or H$\beta$ and no more than 15\% of the flux at the
wavelength of the [\ion{O}{2}]~$\lambda$3727 line. The residual
spectrum is shown in Figure~\ref{fig:spectra} with the emission lines
identified.  It shows the broad H$\alpha$ line very clearly, as well
as an unambiguously broad H$\beta$ line. A hint of a broad H$\gamma$
line is discernible as well. It is worth emphasizing that a careful
subtraction of the starlight is needed to isolate the emission lines
in the vicinity of H$\beta$ since the starlight spectrum has a rather
rich absorption-line structure in this region.  It can easily hide or
distort the appearance of the H$\beta$ line, as well as the nearby
[\ion{O}{3}]~$\lambda$4959 line. Our approach to subtracting the
starlight from the Mt. Hopkins spectrum was somewhat different, since
the flux scale of this spectrum is not calibrated. We used as a
template the spectrum of the S0 galaxy M42 (NGC~4472), observed during
the CfA redshift with the same setup as and within 7 weeks of
NGC~3065. We subtracted the template spectrum from that of NGC~3065,
after normalizing the continuum around H$\alpha$ to unity in both
spectra. This procedure resulted in the removal of all stellar
absorption features from the spectrum of NGC~3065.

A lower limit to the equivalent width of the H$\alpha$ line, {\it
relative to the non-stellar continuum} is $EW > 170$~\AA. This limit
is comparable to the equivalent widths of other LINERs with broad
H$\alpha$ lines (e.g., NGC~4579, NGC~4450, NGC~4203, M81), which are
in the range 140--530~\AA. For comparison, the equivalent widths of
the H$\alpha$ lines of well-known Seyfert galaxies \citep[e.g.,
NGC~4151, NGC~5548, Mkn~6, Mkn~841; measured from the spectra
of][]{eh93} are somewhat larger, falling in the range of 500-600~\AA.

\medskip
\centerline{\psfig{file=figure2.ps,width=2.5in}}
\figcaption[figure2.ps] {Velocity-aligned profiles of the Balmer lines
of NGC~3065 with the lines, observatories, and dates labelled. The
vertical scale of the Mt. Hopkins spectrum is in normalized counts,
while in all other cases it represents $f_{\nu}$ in mJy. The solid
line in the top panel is a fit of a disk model to the broad line
profile. The model is described in \S4 of the text. In the third panel
from the top we superpose the 1980 H$\alpha$ spectrum on the 2000
H$\alpha$ spectrum for comparison (the former spectrum has been scaled
to match the [\ion{N}{2}]~$\lambda$6584 strength of the latter; see \S3 of
the text for more details).
\label{fig:profiles}}
\bigskip

In Table~1 we list the relative emission-line intensities measured
from the spectra after subtracting the continuum. They have been
corrected for Galactic reddening using $E(B-V)=0.067$
\citep*{sfd98}. We also list the widths of the broad Balmer lines as
well as the widths of the narrow lines included in the KPNO spectrum
(unfortunately, the narrow lines in the MDM spectrum are not
resolved). If we naively fit the broad Balmer lines with single
Gaussian models, we find that their centroids are blueshifted by about
450~km~s$^{-1}$ from the reference frame defined by the narrow lines.
However, we also note that at least the H$\alpha$ line shows a red
asymmetry: the red wing extends about 2,000~km~s$^{-1}$ further from
the narrow component of the line than the blue wing does. To
illustrate the properties of the broad-line profiles we plot H$\alpha$
and H$\beta$ spectra on a common velocity scale in
Figure~\ref{fig:profiles}. This illustration brings out the features
of the Balmer line profiles: a blueshifted shoulder and red wing that
is more extended than the blue wing. It also shows that the broad
H$\alpha$ line was absent in the 1980 spectrum, which means that it
must have appeared some time in the past 20 years. To illustrate the
difference between the 1980 and 2000 H$\alpha$ spectra more clearly,
we superpose the two in the third panel of
Figure~\ref{fig:profiles}. The 1980 spectrum has been scaled to match
the [\ion{N}{2}]~$\lambda$6584 strength of the 2000 spectrum. The
difference between the two is easily discernible, especially on the
blue side of the broad H$\alpha$ line, where the intensity in the 2000
spectrum is 4 times higher than what would be consistent with the
noise of the 1980 spectrum.

\medskip
\vskip -3.3in
\hskip -2.6in\leftline{\psfig{file=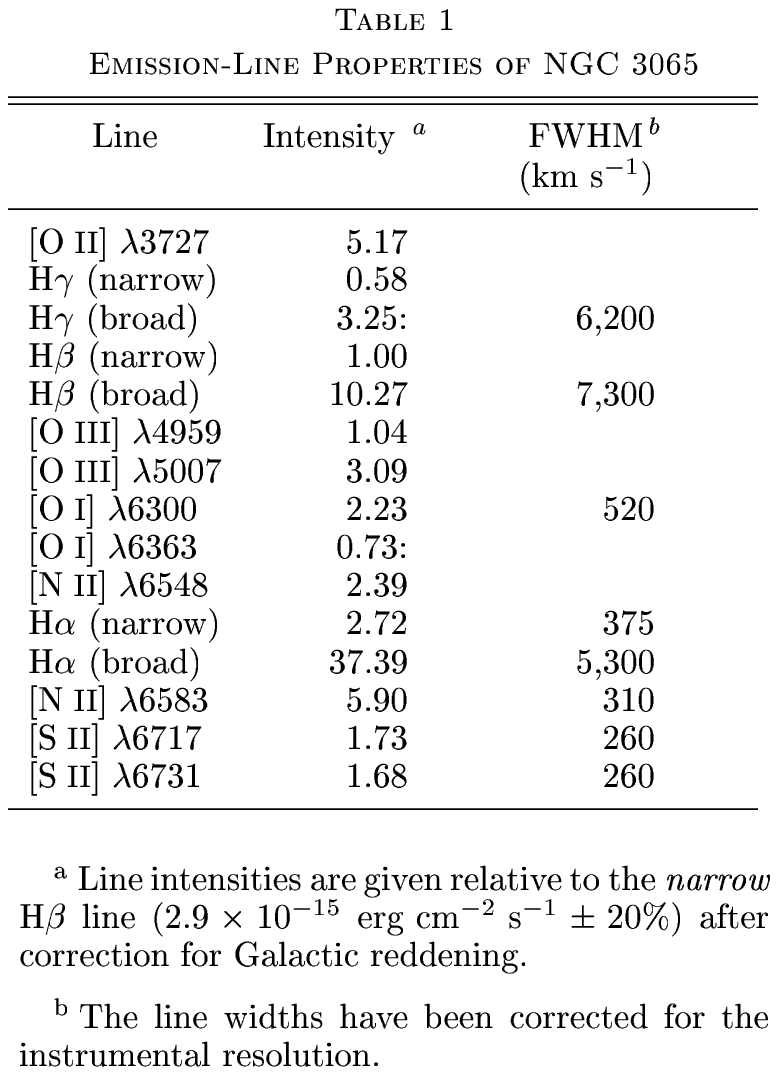,width=8.4in,rheight=7.6in,rwidth=2in}}
\bigskip

To asses whether the normalization procedure we adopted above is
fair, we examined our 2-dimensional spectra of NGC~3065 to search for
extended [\ion{N}{2}]~$\lambda$6584 emission. If the
[\ion{N}{2}]~$\lambda$6584 flux were distributed along the slit, then
the flux collected in the small apertures we used to extract the 2000
spectra would be considerably less than the flux collected in the
larger aperture used in the 1980 observation. Our inspection showed
the [\ion{N}{2}]~$\lambda$6584 source to be unresolved. Thus the
relative strengths of [\ion{N}{2}]~$\lambda$6584 and broad H$\alpha$
are not affected by any missing [\ion{N}{2}]~$\lambda$6584 flux and
the above normalization procedure is valid.

\section{Results, Discussion, and Speculation}

The relative intensities of its narrow lines make NGC~3065 a
LINER. The oxygen line ratios,
[\ion{O}{2}]~$\lambda$3727/[\ion{O}{3}]~$\lambda5007=1.6$ and
[\ion{O}{1}]~$\lambda$6300/[\ion{O}{3}]~$\lambda5007=0.7$, satisfy
Heckman's original definition of the class. Other line ratios, such as
[\ion{O}{3}]~$\lambda$5007/H$\beta = 3.3$,
[\ion{N}{2}]~$\lambda$6583/H$\alpha = 2.2$,
[\ion{O}{1}]~$\lambda$6300/H$\alpha = 0.8$, and
[\ion{S}{2}]~$\lambda\lambda$6717,6731/H$\alpha = 1.3$, place NGC~3065
in the regions of the diagnostic ratio diagrams occupied by LINERs,
albeit close to the boundary with Seyferts \citep*[see, for
example,][]{hfs97}. The widths of the narrow forbidden lines appear to
follow a trend with critical density: lines of higher critical
density are broader than lines of lower critical density. Such a trend
is often observed in LINERs and it has been interpreted as an
indication that the narrow-line emitting gas is stratified in density
\citep{f85}. The presence of broad Balmer lines in the spectrum of
NGC~3065 establishes it as an AGN beyond doubt, and confirms the
suggestion of \citet{i98} based on the X-ray properties.

Finding broad Balmer lines in the spectra of LINERs is not unusual.
For example, M81 has been known for quite some time to have broad
Balmer lines \citep{pt81,fs88}. More recently, broad Balmer lines have
been found in several LINERs with the {\it Hubble Space Telescope
(HST)} which can obtain spectra through very small apertures that
exclude the contaminating starlight very effectively. Examples include
NGC~4203 \citep{s00}, NGC~4450 \citep{h00}, and NGC~4579
\citep{b00}. The sudden appearance or dramatic variability of broad
Balmer lines in LINERs is not unheard of either: it has been observed
in at least two other cases so far, NGC~1097 \citep{sbw93} and M81
\citep{b96}.  It is also possible that the broad H$\alpha$ lines of
NGC~4203, NGC~4450, and NGC~4579 varied dramatically between the early
observations from the mid-1980s \citep{hfsp97} and the later {\it
HST} observations from the late 1990s, although one cannot be
confident in view of the available data.

The broad Balmer lines of LINERs often have double-peaked profiles,
which are characteristic of rotation and suggest an origin in an
accretion disk around a supermassive black hole (see above
references). This underscores an intimate connection between LINERs
and another class of double-peaked emission line AGNs, the broad-line
radio galaxies \citep[hereafter BLRGs;][]{eh94}, which is bolstered by
other similarities between the two classes of object.  In particular,
the relative strengths of the narrow emission lines of BLRGs with
double-peaked emission lines approach those of LINERs \citep*[the
prototype, Arp~102B is a certified LINER;][]{ssk83}. Moreover, in
Pictor~A, whose relative narrow-line strengths are close to LINER-like
\citep{f85}, the double-peaked Balmer lines appeared abruptly in the
mid- to late-1980s \citep{he94,sul95}.

Although the broad Balmer lines of NGC~3065 are not double peaked, the
fact that their red wing extends further than the blue wing is
reminiscent of gravitational redshift of photons originating in the
inner part of an accretion disk. We thus speculate that the broad
Balmer lines of NGC~3065 originate in the outer parts of an accretion
disk around a supermassive black hole.  We emphasize that this is by
no means a unique explanation for the origin of the broad Balmer
lines; it is merely inspired by the profiles of the Balmer lines of
other similar objects. To assess the plausibility of this hypothesis
we have tried to fit their profiles with the disk model developed by
\citet*{chf89} and \citet{ch89}. The result of this exercise is
superposed on the observed H$\alpha$ profile shown in the top panel of
Figure~\ref{fig:profiles}. According to the adopted model, the axis of
the disk is inclined at an angle of $50^{\circ}$ to the line of sight
and the line-emitting region is between radii of 900 and
$100,000~GM/c^2$, where $M$ is the mass of the black hole. The
emissivity is a broken power law with radius: $\epsilon\propto
r^{-q}$, with $q=1.7$ for $r<10,000~GM/c^2$ and $q=3$ elsewhere.  In
this context it is noteworthy that the profiles of emission lines
coming from an accretion disk need not be double-peaked: if the ratio
of the inner-to-outer radius of the line-emitting part of the disk is
large or the disk is close to face on, the two peaks get close enough
together that they merge and the profile appears flat-topped or
single-peaked \citep{e99,c97}. Other combinations of model parameters
may be able to produce equally good fits. We have not explored the
parameter space because we are only focusing on the plausibility of
this interpretation here.  Also, more sophisticated disk models which
include an eccentricity \citep{e95} or a spiral wave \citep{g99} may
be able to reproduce the observed H$\alpha$ profile even better, but
such detailed modeling is well outside the scope of this paper.

To further explore the similarity of NGC~3065 with LINERs and BLRGs
with double-peaked Balmer lines we have estimated the mass of its
central black hole and the corresponding Eddington luminosity.  To
estimate the black hole mass we used the recently established
correlation between it and the stellar velocity dispersion in the host
galaxy \citep{fm00,g00}. The stellar velocity dispersion of
$173\pm16~{\rm km~s}^{-1}$ reported by \citet{td81} thus yielded
$M_{\bullet}=(9\pm4)\times 10^7~{\rm M}_{\odot}$, where the error bar
reflects not only the uncertainty in the velocity dispersion but also
uncertainties in the parameters describing the correlation between the
velocity dispersion and the black hole mass \citep[see,][]{fm00,g00}.
As a check, we also estimated the black hole mass based on the
correlation between it and the blue luminosity of the bulge of the
host galaxy \cite[see the latest version in][]{k00} and using the
bulge-disk decomposition of \citet{k77}, obtaining
$M_{\bullet}=2\times 10^8~{\rm M}_{\odot}$. This value is almost a
factor of 2 higher than that obtained with the previous method, which
is very likely a result of the large scatter about the mean trend
between the black hole mass and the bulge luminosity.  Because of this
large dispersion, we prefer the black hole mass inferred from the
stellar velocity dispersion.  The implied Eddington luminosity is
$L_{\rm Edd}=1.3\times 10^{46}~{\rm erg~s^{-1}}$. If we take the
bolometric accretion luminosity of NGC~3065 to be 10 times larger than
the observed 2--10~keV X-ray luminosity \citep[see, for
example,][]{h99}, we find an Eddington ratio of $L_{\rm bol}/L_{\rm
Edd}\approx 2\times 10^{-4}$, which indicates a very low relative
accretion rate, a common feature of LINERs \citep[see][]{h99}.  At
such a low accretion rate the inner accretion disk is likely to be
``advection dominated'' \citep[an ADAF or ion torus;][]{ny94,ny95,r82}
and to form a vertically extended structure that can illuminate the
outer, thin disk effectively. Thus, it could power the observed
broad-line emission \citep[cf,][]{ch89}. This geometrical requirement
may very well be the reason why disk-like emission lines are
preferentially found in AGNs with very low accretion rates relative to
the Eddington rate. In fact, if we are to associate the broad Balmer
lines of NGC~3065 with emission from an accretion disk, then external
illumination of the disk is needed in order to power the line
emission. This is because the H$\alpha$ luminosity is $3\times
10^{40}~{\rm erg~s^{-1}}$ while the viscous power output of the
line-emitting portion of the disk \citep[calculated following][]{eh94}
is only $4\times 10^{39}~{\rm erg~s^{-1}}$. Yet another appealing
feature of ADAFs in LINERs is the fact that their hard spectral energy
distribution, which lacks a ``UV bump,'' when combined with a low
ionization parameter, can explain the relative strengths of the narrow
emission lines \citep{hs83,fn83}.

The reason for the recent emergence of the broad Balmer lines in
NGC~3065, as well as in similarly behaved objects, remains an open
question. One possibility is that the emission lines come from a
transient accretion disk which formed from the debris released by the
tidal disruption of a star by the black hole \citep[cf,
NGC~1097;][]{e95,s95}. Another is a change in the structure of the
inner accretion disk associated with a change in the accretion rate,
i.e., a transformation from a thin disk to an ADAF
\citep{s97}. Perhaps the long-term variations of the Balmer-line
profiles will provide clues to their origin. The variations may show
evidence for dynamical phenomena \citep[e.g., spiral waves in the
disk;][]{g99}, which may cause fluctuations in the accretion rate.
Alternatively, the variations may show evidence for a disk geometry
that can be related to its formation process \citep[e.g., an eccentric
disk formed from tidal debris;][]{e95}.  We will continue to monitor
the broad Balmer lines of NGC~3065 in an effort to uncover their
cause.

\acknowledgements

We thank J. Huchra for sending us the spectra of NGC~3065 and NGC~4472
obtained during the CfA Redshift Survey. We are grateful to Luis Ho,
Aaron Barth, and the anonymous referee for their insightful comments
and suggestions.  In our investigation we have made use of the
NASA/IPAC Extragalactic Database (NED) which is operated by the Jet
Propulsion Laboratory, California Institute of Technology, under
contract with the National Aeronautics and Space Administration.


\begin{thebibliography}{}

\bibitem[\protect\citeauthoryear{Barth et al.}{2000}]{b00} 
Barth, A. J., Ho, L. C., Filippenko, A. V., Rix, H.-W., \& Sargent,
W. L. W. 2000, \apj, in press (astro-ph/0008273)

\bibitem[\protect\citeauthoryear{Bower et al.}{1996}]{b96} 
Bower, G. A., Wilson, A. S., Heckman, T. M. \& Richstone, D. O.
1996 \aj, 111, 1901

\bibitem[\protect\citeauthoryear{Burbidge \& Burbidge}{1965}]{bb65} 
Burbidge, E. M. \& Burbidge, G. R. 1965, \apj, 142, 634


\bibitem[\protect\citeauthoryear{Chen \& Halpern}{1989}]{ch89}
Chen, K. \& Halpern, J. P. 1989, \apj, 344, 115

\bibitem[\protect\citeauthoryear{Chen, Halpern, \& Filippenko}
{Chen et al.}{1989}]{chf89}
Chen, K., Halpern, J. P., \& Filippenko, A. V. 1989, \apj, 339, 742

\bibitem[\protect\citeauthoryear{Condon et al.}{1998}]{c98} 
Condon, J. J., Cotton, W. D., Greisen, E. W., Yin, Q. F.,
Perley, R. A., \& Taylor, G. B. 1998, \apj, 115, 1693

\bibitem[\protect\citeauthoryear{Corbin}{1997}]{c97}
Corbin, M. R  1997, \apj, 485, 517

\bibitem[\protect\citeauthoryear{Eracleous}{1999}]{e99}
Eracleous, M. 1999, in ``Structure and Kinematics of Quasar Broad Line
Regions,'' ASP Conference Series, Vol. 175. eds. C. M. Gaskell, et al.
(San Francisco: ASP), 163

\bibitem[\protect\citeauthoryear{Eracleous \& Halpern}{1993}]{eh93} 
Eracleous, M. \& Halpern, J. P. 1993, \apj, 409, 584

\bibitem[\protect\citeauthoryear{Eracleous \& Halpern}{1994}]{eh94} 
Eracleous, M. \& Halpern, J. P. 1994, \apjs, 90, 1

\bibitem[\protect\citeauthoryear{Eracleous et al.}{1995}]{e95} 
Eracleous, M., Livio, M., Halpern, J. P., \& Storchi-Bergmann,
T. 1995, \apj, 438, 610

\bibitem[\protect\citeauthoryear{Fabbiano, Kim, \& Trinchieri}
{Fabbiano et al.}{1992}]{fkt92} 
Fabbiano, G., Kim, D.-W., \& Trinchieri, G. 1992, \apjs, 80, 531

\bibitem[\protect\citeauthoryear{Ferland \& Netzer}{1983}]{fn83} 
Ferland, G. J. \& Netzer, H. 1983, \apj, 264, 105

\bibitem[\protect\citeauthoryear{Ferrarese \& Merrit}{2000}]{fm00} 
Ferrarese, L. \& Merrit, D. 2000, \apj, 539, L9

\bibitem[\protect\citeauthoryear{Filippenko}{1996}]{f96} 
Filippenko, A. V. 1996, ``The Physics of LINERs in View of Recent
Observations,'' ASP Conference Series, Vol. 103; eds. M. Eracleous et
al. (San Francisco: ASP), 17

\bibitem[\protect\citeauthoryear{Filippenko}{1985}]{f85} 
Filippenko, A. V. 1985, \apj, 289, 475

\bibitem[\protect\citeauthoryear{Filippenko \& Sargent}{1988}]{fs88} 
Filippenko, A. V. \& Sargent, W. L. W. 1988, \apj, 324, 134

\bibitem[\protect\citeauthoryear{Gebhardt et al.}{2000}]{g00} 
Gebhardt, K., et al. 2000, \apj, 539, L13

\bibitem[\protect\citeauthoryear{Gilbert et al.}{1999}]{g99}
Gilbert, A. M., Eracleous, M., Filippenko, A. V., \& Halpern
J. P. 2000, in ``Structure and Kinematics of Quasar Broad Line Regions,
ASP Conference Series,'' Vol. 175. eds. C. M. Gaskell, et al.  (San
Francisco: ASP), 189

\bibitem[\protect\citeauthoryear{Halpern \& Eracleous}{1994}]{he94}
Halpern, J. P. \& Eracleous, M. 1994, \apj, 433, L17

\bibitem[\protect\citeauthoryear{Halpern \& Steiner}{1983}]{hs83}
Halpern, J. P. \& Steiner, J. E. 1983, \apj, 269, L37

\bibitem[\protect\citeauthoryear{Heckman}{1980}]{h80} 
Heckman, T. M. 1980, A\&A, 87, 152

\bibitem[\protect\citeauthoryear{Ho}{1999}]{h99} 
Ho, L. C. 1999, \apj, 516, 672

\bibitem[\protect\citeauthoryear{Ho et al.}{1997a}]{hfs97} 
Ho, L. C., Filippenko, A. V., \& Sargent, W. L. W. 1997a, \apjs, 112, 315

\bibitem[\protect\citeauthoryear{Ho et al.}{1997b}]{hfsp97} 
Ho, L. C., Filippenko, A. V., \& Sargent, W. L. W., \& Peng, C. Y. 
1997b, \apjs, 112, 391

\bibitem[\protect\citeauthoryear{Ho et al.}{2000}]{h00} 
Ho, L. C., Rix, H.-W., Shields, J. C., Rudnick, G., McIntosh, D. H.,
Filippenko, A. V., \& Sargent, W. L. W. , \& Eracleous, M. 2000,
\apj, 541, 120

\bibitem[\protect\citeauthoryear{Humason, Mayall, \& Sandage}
{Humason et al.}{1956}]{hms56} 
Humason, M. L., Mayall, N. U., \& Sandage, A. R. 1956, \aj, 61, 97

\bibitem[\protect\citeauthoryear{Iyomoto et al.}{1998}]{i98}
Iyomoto, N., Makishima, K., Matsushita, K., Fukazawa, Y., Tashiro, M.
\& Ohashi, T. 1998, \apj, 503, 168
 
\bibitem[\protect\citeauthoryear{Kormendy}{1977}]{k77} 
Kormendy, J. 1977, \apj, 217, 406

\bibitem[\protect\citeauthoryear{Kormendy}{2000}]{k00} 
Kormendy, J. 2000, in ``Galaxy Disks and Disk Galaxies,''
eds. J. G. Funes \& E. M. Corsini (San Francisco: ASP), in press
(astro-ph/0007401).



\bibitem[\protect\citeauthoryear{Narayan \& Yi}{1994}]{ny94} 
Narayan, R. \& Yi, I. 1994, \apj, 443, L41

\bibitem[\protect\citeauthoryear{Narayan \& Yi}{1995}]{ny95} 
Narayan, R. \& Yi, I. 1995, \apj, 444, 231

\bibitem[\protect\citeauthoryear{Peimbert \& Torres-Peimbert}{1981}]{pt81}
Peimbert, M. \& Torres-Peimbert, S. 1981, \apj, 245, 845.
 

\bibitem[\protect\citeauthoryear{Ptak et al.}{1999}]{p99} 
Ptak, A., Serlemitsos, P., Yaqoob, T., \& Mushotzky, R. F. 1999,
\apjs, 120, 179

\bibitem[\protect\citeauthoryear{Rees et al.}{1982}]{r82} 
Rees, M. J., Begelman, M. C., Blandford, R. D., \& Phinney, E. S. 1982,
Nature, 295, 17


\bibitem[\protect\citeauthoryear{Schlegel, Finkbeiner \& Davis}
{Schlegel et al.}{1998}]{sfd98} 
Schlegel, D., Finkbeiner, D. P., \& Davies, M. 1998, \apj, 500, 525

\bibitem[\protect\citeauthoryear{Shields et al.}{2000}]{s00} 
Shields, J. C., Rudnick, G., Rix, H.-W., Ho, L. C., McIntosh, D. H., 
Filippenko, A. V., \& Sargent, W. L. W. 2000, \apj, 534, L27


\bibitem[\protect\citeauthoryear{Storchi-Bergmann, Baldwin, \&
Wilson,}{Storchi-Bergmann et al.}{1993}]{sbw93}
Storchi-Bergmann, T., Baldwin, J. A., \& Wilson, A. S. 1993, \apj,
410, L11

\bibitem[\protect\citeauthoryear{Storchi-Bergmann et al.}{1995}]{s95}
Storchi-Bergmann, T., Eracleous, M., Livio, M., Wilson, A. S.,
Filippenko, A. V., Halpern, J. P. 1995, \apj, 443, 617

\bibitem[\protect\citeauthoryear{Storchi-Bergmann et al.}{1997}]{s97}
Storchi-Bergmann, T., Eracleous, M., Ruiz, M. T., Livio, M., Wilson,
A. S., \& Filippenko, A. V. 1997, \apj, 489, 87

\bibitem[\protect\citeauthoryear{Stauffer, Schild, \& Keel}{Stauffer 
et al.}{1983}]{ssk83}
Stauffer, J., Schild, R., \& Keel, W. C. 1983, \apj, 270, 465
 
\bibitem[\protect\citeauthoryear{Sulentic et al.}{1995}]{sul95} 
Sulentic, J. W., Marziani, P., Zwitter, T., Calvani, M. 1995, \apj, 438, L1

\bibitem[\protect\citeauthoryear{Terashima et al.}{2000}]{t00}
Terashima, Y., Ho. L. C., Ptak, A. F., Mushotzky, R. F., Serlemitsos,
P., Yaqoob, T., \& Kunieda, H. 2000, \apj, 533, 729

\bibitem[\protect\citeauthoryear{Tonry \& Davis}{1981}]{td81} 
Tonry, J. L. \& Davis, M. 1981, \apj, 246, 666

\bibitem[\protect\citeauthoryear{Tully}{1987}]{t87} 
Tully, R. B. 1987, Nearby Galaxies Catalog (Cambridge: Cambridge
University Press)


\end{thebibliography}
\end{document}